\documentclass[conference]{IEEEtran}
\IEEEoverridecommandlockouts

\usepackage{cite}
\usepackage{amsmath,amssymb,amsfonts}
\usepackage{balance}
\usepackage{algorithmic}
\usepackage{graphicx}
\usepackage{textcomp}
\usepackage{xcolor}
\usepackage{fontawesome}
\usepackage{footnote}
\usepackage{url}
\usepackage{caption}
\usepackage{graphicx}
\usepackage{tcolorbox}
\usepackage{colortbl}
\usepackage{enumitem}
\usepackage{listings}
\usepackage{todonotes}
\usepackage{colortbl}
\usepackage{tcolorbox}
\usepackage{tabularx,ragged2e} 
\usepackage{booktabs}

\definecolor{mygreen}{rgb}{0,0.6,0}
\definecolor{mygray}{rgb}{0.5,0.5,0.5}
\definecolor{mymauve}{rgb}{0.58,0,0.82}
\definecolor{dkgreen}{rgb}{0,0.6,0}
\definecolor{gray}{rgb}{0.5,0.5,0.5}
\definecolor{mauve}{rgb}{0.58,0,0.82}
\definecolor{gray}{rgb}{0.4,0.4,0.4}
\definecolor{darkblue}{rgb}{0.0,0.0,0.6}
\definecolor{lightblue}{rgb}{0.0,0.0,0.9}
\definecolor{cyan}{rgb}{0.0,0.6,0.6}
\definecolor{darkred}{rgb}{0.6,0.0,0.0}
\definecolor{gray50}{gray}{.5}
\definecolor{gray40}{gray}{.6}
\definecolor{gray30}{gray}{.7}
\definecolor{gray20}{gray}{.8}
\definecolor{gray10}{gray}{.9}
\definecolor{gray05}{gray}{.95}

\newcommand{\revised}[1]{\textcolor{black}{#1}}

\newcommand{\resquestion}[2]{ %
	\vspace{5pt} %
	\noindent\fcolorbox{black}{blue!05}{%
		\parbox{0.97\linewidth}{%
			\textbf{RQ$_{#1}$.} \emph{#2} 
		}%
	}%
	\vspace{5pt} %
}%

\def\BibTeX{{\rm B\kern-.05em{\sc i\kern-.025em b}\kern-.08em
    T\kern-.1667em\lower.7ex\hbox{E}\kern-.125emX}}
    
\begin{document}

\title{Test Code Refactoring Unveiled: Where and How Does It Affect Test Code Quality and Effectiveness?}

\author{\IEEEauthorblockN{Luana Martins\IEEEauthorrefmark{1}, Valeria Pontillo\IEEEauthorrefmark{2}, Heitor Costa\IEEEauthorrefmark{3}, Filomena Ferrucci\IEEEauthorrefmark{2}, Fabio Palomba\IEEEauthorrefmark{2}, Ivan Machado\IEEEauthorrefmark{1}}

\IEEEauthorblockA{martins.luana@ufba.br, vpontillo@unisa.it, heitor@ufla.br, fferrucci@unisa.it, fpalomba@unisa.it, ivan.machado@ufba.br}

\IEEEauthorblockA{
    \IEEEauthorrefmark{1}Federal University of Bahia, Salvador, Brazil
    }
    
\IEEEauthorblockA{
    \IEEEauthorrefmark{2}Software Engineering (SeSa) Lab --- University of Salerno, Fisciano, Italy
    }

\IEEEauthorblockA{
    \IEEEauthorrefmark{3}Federal University of Lavras, Lavras, Brazil 
    }
}

\maketitle
\thispagestyle{plain}
\pagestyle{plain}

\begin{abstract}
\textit{Context.} Refactoring has been widely investigated in the past in relation to production code quality, yet still little is known on how developers apply refactoring on test code. Specifically, there is still a lack of investigation into how developers typically refactor test code and its effects on test code quality and effectiveness.
\textit{Objective.} This paper presents a research agenda aimed to bridge this gap of knowledge by investigating (1) whether test refactoring actually targets test classes affected by quality and effectiveness concerns and (2) the extent to which refactoring contributes to the improvement of test code quality and effectiveness. \textit{Method.} We plan to conduct an exploratory mining software repository study to collect test refactoring data of \revised{open-source \textsc{Java} projects from \textsc{GitHub}} and statistically analyze them in combination with quality metrics, test smells, and code/mutation coverage indicators. Furthermore, we will measure how refactoring operations impact the quality and effectiveness of test code. 

\end{abstract}


\section{Introduction}
\label{section:Introduction}
Refactoring is an engineered approach that allows developers to improve the quality of source code without affecting its external behavior \cite{Fowler18}. 
Over the last decades, researchers have been proposing automated refactoring recommenders \cite{bavota2014recommending} and investigated how refactoring relates to code quality \cite{azeem2019machine,de2018systematic,DuBois2004_Refactoring_to_improve_metrics, Chavez2017_Refactoring_and_internal_attr}. \revised{In particular, researchers identified both benefits and drawbacks of its application \cite{al2015identifying, baqais2020automatic, lacerda2020code}, finding that, while refactoring is theoretically associated with modifications that do not affect the external behavior of source code, it may possibly induce defects \cite{di2020relationship, bavota2012does, ferreira2018}, vulnerabilities \cite{iannone2023rubbing}, or even code smells \cite{tufano2015and}. These drawbacks are mainly due to refactoring activities performed manually without the support of automated tools and interleaved with other code changes \cite{murphy2007don}. Our research is motivated by these previous works. On the one hand, most previous studies focused on the refactoring of production code and, for this reason, we argue that there is a lack of investigations into \emph{how refactoring is applied to test code}. On the other hand, we do not know if similar effects observed in previous work may arise with test refactoring, i.e., it may have some impact on both test quality and effectiveness, for instance, in cases where refactoring actions target the logic of a test case. Hence, we point out a \emph{limited knowledge on the effects of refactoring} on both test quality and effectiveness.}

An improved understanding of test refactoring would have a number of potential benefits for research and practice. In the first place, test cases represent a crucial asset for software dependability: developer's productivity is partly dependent on the quality of test cases \cite{micco2017state}, as these help practitioners to decide on whether to merge pull requests or deploy the system \cite{grano2020pizza}. As such, analyzing how refactoring affects test cases may have a significant impact on practice. Secondly, researchers have been showing that the design of test code is approached in a substantially different way with respect to traditional development \cite{meszaros2007xunit}. Indeed, the test code must often interact with external systems, databases, or APIs to set up test environments and verify the system's behavior \cite{Meszaros2003_Manifesto}. As a consequence, test code may suffer from different issues that, in turn, would require different refactoring operations \cite{Guerra2007RefactoringSafely}.

\revised{For these reasons, new refactoring practices have been proposed with the aim of dealing with quality or effectiveness concerns \cite{Deursen01_Refactoring, Meszaros2003_Manifesto, Guerra2007RefactoringSafely}. While those refactoring practices were the target of some previous investigations, researchers limited their focus to how refactoring may influence test smells, i.e., symptoms of poor test code quality \cite{Soares2020_Refactoring_Smells,Soares22_JUnit5,Peruma2020_Refactoring_Tests_Prod}, hence not providing comprehensive analysis into the \emph{nature} and \emph{effects} of test refactoring.
More specifically, we highlight a lack of knowledge on (1) whether developers apply test refactoring operations on test classes that are actually affected by quality or effectiveness concerns, as it is supposed to be given the definition of refactoring; and (2) what is the effect of refactoring on both quality and effectiveness of test cases.}

This paper aims at addressing this gap of knowledge by proposing an \emph{exploratory empirical study}. We first plan to collect test refactoring data from the change history of \revised{open-source \textsc{Java} projects from \textsc{GitHub}}
~and combine them with data coming from automated instruments able to profile test code from the perspective of quality metrics, test smells, and code/mutation coverage information. Afterward, we plan to apply statistical analyses to address three main research goals targeting (1) whether test classes with a low level of quality, in terms of test smells and code metrics, are associated with more test refactoring, (2) whether a low level of effectiveness, in terms of mutation coverage and code coverage, is associated with more test refactoring, and (3) to what extent the removal of test smells improve the test code quality and effectiveness.

\revised{Our findings might benefit researchers and practitioners under multiple perspectives. In the first place, our research may reveal insights into the refactoring types that may deteriorate test code quality and effectiveness. Such information would be relevant for researchers in both the fields of refactoring and testing, as it may lead them to (1) extend the knowledge on the best and bad practices to properly apply test refactoring; (2) devise novel test refactoring approaches which are aware of the possible side effects of refactoring, e.g., we may envision multi-objective search-based refactoring approaches that may optimize refactoring recommendations based on both quality and effectiveness attributes; and (3) design novel recommendation systems that may support developers in understanding how a refactoring would impact different test code properties. The results would also be useful to practitioners, who may have additional proof of the side effects of refactoring, hence possibly being stimulated further on the need to employ automated refactoring tools. In the second place, our findings may indicate the nature of the test cases more likely to be subject to refactoring operations. Researchers might use this information to define refactoring recommenders and refactoring prioritization approaches, while practitioners may acquire the awareness of their actions.
}

\section{Related Work}
\label{section:Related_Work}
The current literature can be distinguished based on the type of empirical studies conducted. First, several studies analyzed change history information to extract knowledge about test smells and their impact. Spadini et al. \cite{Spadini2018_CodeQuality} investigated ten open-source projects to find a relation between six test smells and the change and defect-proneness of both test and production code, finding that smelly JUnit tests are more change-prone and defect-prone than non-smelly ones. In addition, they found that production code is typically more defect-prone when tested by smelly tests. \revised{As such, the authors did not target test code refactoring, hence not assessing how the seemingly test code quality improvement actions performed by developers affect test code quality and effectiveness, i.e., the authors looked exactly in the opposite direction of our paper, focusing on how bad practices affect test code quality.} 

Wu et al. \cite{Haitao2022_ImproveQuality} explored the impact of eliminating test smells on the production code quality of ten open-source projects. 
\revised{In this respect, there are two key points that make our investigation novel: (1) test smell removal does not imply the application of refactoring: a previous empirical study \cite{Kim2021_Secret_Life} indeed showed that 83\% of test smell removal activities are due to feature maintenance actions, i.e., our work can therefore further the knowledge on how developers apply test code refactoring; (2) the authors worked, also in this case, in the opposite direction as our work, focusing on the effects of test smells on code quality rather than analyzing the impact of test code refactoring actions. As such, our work extends the current knowledge by assessing how test refactoring is applied and what is its impact on multiple aspects of test code, such as quality and effectiveness.} 

Peruma et al. \cite{Peruma2020_Refactoring_Tests_Prod} investigated the relationship between refactoring changes and their effect on test smells. The authors used \textsc{Refactoring Miner} \cite{RefactoringMiner2022} to detect refactoring operations and the \textsc{tsDetect} tool \cite{Peruma2020tsdetect} to identify the test smells from unit test files of 250 open-source Android Apps. Results showed that refactoring operations in test and non-test files differ, and the refactorings co-occur with test smells. \revised{With respect to the work by Peruma et al. \cite{Peruma2020tsdetect}, we do not limit ourselves to the analysis of test smells, but also consider additional indicators of test code quality and effectiveness: in this sense, ours will represent a more comprehensive analysis of the role of test refactoring. Second, we assess the actual effects of test refactoring on test code quality and effectiveness, providing insights into how various test refactoring types may support the evolutionary activities of developers.} 

A second line of research is represented by qualitative studies targeting the developer's perception of test refactoring. Damasceno et al. \cite{Damasceno2023_Analyzing_Refactorings} investigated the impact of test smell refactoring on internal quality attributes, \revised{reporting some insights that may potentially be in line with the results of our study, e.g., they let emerge the impact of test smell refactoring on internal quality attributes. In the first place, the authors specifically focused on the refactoring of test smells, while our work targets test code refactoring from a more general perspective, attempting to assess the extent to which this is applied to classes suggesting the presence of quality or effectiveness concerns. Secondly, the results of our work may possibly provide evidence-based, complementary insights with respect to what the authors found out in their qualitative study. Third, our work has a broader scope and, indeed, it also targets the effectiveness side of the problem.} 
Soares et al. \cite{Soares2020_Refactoring_Smells} investigated how developers refactor test code to eliminate test smells. The authors surveyed 73 open-source developers and submitted 50 pull requests to assess developers’ preferences and motivation while refactoring the test code. The results showed that developers preferred the refactored test code for most test smells. In another work, Soares et al. \cite{Soares22_JUnit5} investigated whether the JUnit 5 features help refactor test code to remove test smells. They conducted a mixed-method study to analyze the usage of the testing framework features in 485 popular Java open-source projects, identifying the features helpful for test smell removal and proposing novel refactorings to fix test smells. \revised{Also in this case, the authors focused on the refactoring of test smells, while our study has a broader scope. In addition, while we do not plan to conduct surveys or interviews---this is part of our future research agenda---, we will extend the current body of knowledge by assessing whether test code quality and effectiveness indicators may trigger refactoring activities, other than providing a comprehensive overview of how test refactoring relates to branch and mutation coverage, which is a premiere of our study.}
\section{Research Questions and Objectives}
\label{section:Research_Design}

The \textit{goal} of the empirical study is to analyze the test refactoring operations performed by developers over the history of software projects, with the \textit{purpose} of understanding (1) whether low-quality test classes, in terms of structural metrics and test smells, provide indications on which test classes are more likely of being refactored, (2) whether test classes with low effectiveness, in terms of code coverage and mutation coverage, provide indications on which test classes are more likely of being refactored, and (3) as a consequence, to what extent test refactoring operations are effective in improving quality and effectiveness of test classes. In other terms, we are first interested in assessing the \textbf{quantity} of test refactoring operations performed on classes exhibiting test code quality and effectiveness issues and, in the second place, the \textbf{quality} of the test refactoring operations applied in terms of improvements provided to test code quality and effectiveness. The \textit{perspective} is of both researchers and practitioners who are interested in understanding the relationship and effects of test refactoring operations on the quality and effectiveness of test classes.

More specifically, our empirical investigation will first aim at addressing the following research questions (\textbf{RQ}s):

\resquestion{1}{Are test refactoring operations performed on test classes having a low level of quality, as indicated by quality metrics and test smell detectors?}

\resquestion{2}{Are test refactoring operations performed on test classes having a low level of effectiveness, as indicated by code and mutation coverage?}

Through \textbf{RQ$_{1}$} and \textbf{RQ$_{2}$}, we aim to address the first objective of the study, hence understanding whether the low quality and effectiveness of test classes are associated with more test refactoring operations. The results of these two research questions might have multiple implications for software maintenance, evolution, and testing researchers. An improved understanding of these aspects may indeed inform researchers on the characteristics of the test suites that trigger more refactoring operations, \revised{possibly informing researchers on (1) the factors that are associated with test refactoring and (2) the design of novel or improved instruments to better support developers in their activities.} For instance, should we discover that test refactoring is not frequently applied on test classes exhibiting test smells, this would imply that further research should be conducted on the motivations leading developers to refactor test code, other than to how test smell detectors should be designed to ease the application of refactoring operations. 

Upon completion of this investigation, we will further elaborate on the impact of test refactoring, addressing the following research questions:

\resquestion{3}{What is the effect of test refactoring on test code quality, as indicated by quality metrics and test smell detectors?}

\resquestion{4}{What is the effect of test refactoring on test code effectiveness, as indicated by code and mutation coverage?}

Through \textbf{RQ$_{3}$} and \textbf{RQ$_{4}$}, we aim to extend the current knowledge on the impact of test refactoring, assessing whether the test code quality and effectiveness increase, decrease or remain the same after the application of test refactoring operations. It is worth to remark that addressing these two research questions would be important independently from the results obtained by \textbf{RQ$_{1}$} and \textbf{RQ$_{2}$}. Indeed, regardless of the amount of refactoring operations performed on test classes exhibiting quality or effectiveness concerns, it would still be possible that the specific refactoring actions targeting those classes have an impact. \revised{To make our argumentation more practical, consider the case of the \emph{Extract Method} refactoring, whose suboptimal implementation may potentially affect test code effectiveness. Given a verbose test method with several steps and assertions, the refactoring enables the extraction of multiple test methods, which are supposed to be more cohesive and focused on the verification of specific conditions of production methods. However, if developers do not appropriately perform such an extraction, this would potentially change the logic of the test and be detrimental to test effectiveness. For instance, consider test T, which verifies two branches, B1 and B2, of the production method M. In this case, an Extract Method operation is supposed to split T so that the resulting tests T1 and T2 target B1 and B2 individually. However, should there be a logical relation between B1 and B2, T2 will still need to pass through T1 to ensure that the logical relation is still met: a suboptimal refactoring may overlook this requirement, possibly not embedding in T2 the statements required to reach B1. As a result, this operation would affect the overall level of coverage of the production code.}

As such, \textbf{RQ$_{3}$} and \textbf{RQ$_{4}$} provide an orthogonal view on the matter. Also in this case, the outcome of our investigation may lead to implications for research and practice. First, our findings may help researchers measure the actual, practical impact of test refactoring---this may drive considerations on how future research efforts should be prioritized, e.g., by favoring more research on impactful refactoring operations. Second, our results may increase the practitioner's awareness of test refactoring, possibly increasing its application in practice. 

To design and report our empirical study, we will follow the empirical software engineering guidelines by Wohlin et al. \cite{Wohlin2012_Experimentation} other than the ACM/SIGSOFT Empirical Standards.\footnote{Available at: \url{https://github.com/acmsigsoft/EmpiricalStandards}}

\section{Experimental Plan}
\label{section:Experimental_Plan}
This section reports the research method that we plan to apply to address our \textbf{RQs}. 

\subsection{Context of the study}
\label{subsection:Subject_Projects}
The \emph{context} of our investigation will be composed of (i) empirical study variables, i.e., the independent and dependent variables that we will statistically analyze, and (ii) software systems, i.e., the projects that will be mined to collect the data required to address our research objectives.

\begin{table*}[!t]
\centering \footnotesize
\caption{Description of quality metrics as detected by \textsc{VITRuM} \cite{Pecorelli2020_Vitrum}}
\setlength{\tabcolsep}{0.2cm}
\label{table:metricsDescription}

\begin{tabular}{l|l|p{11cm}} \toprule
\rowcolor{gray20} \multicolumn{1}{c}{\textbf{Acronym}} & \multicolumn{1}{c}{\textbf{Quality Metrics}} & \multicolumn{1}{c}{\textbf{Description}} \\ \midrule
\rowcolor{gray10} LOC & Number of Lines	& Counts the number of lines \\
NOM & Number of Methods	&  Counts the number of methods \\
\rowcolor{gray10} WMC & Weight Method Class	& Counts the number of branch instructions in a class \\
RFC	& Response for a Class	& Counts the number of method invocations in a class \\
\rowcolor{gray10}AD	& Assertion density	& Percentage of assert statements with respect to the total number of statements in a test class \\
MUT & Mutation Coverage &  Percentage of mutated statements in the production class that is covered by the test\\
\rowcolor{gray10}LCOV & Line coverage	& Lines exercised by the test\\
BCOV	& Branch coverage	& Branches exercised by the test\\ \midrule
\end{tabular}
\end{table*}

\begin{table*}[!t]
\centering \footnotesize
\caption{Description of test smells as detected by \textsc{tsDetect} \cite{Peruma2020tsdetect}}
\setlength{\tabcolsep}{0.2cm}
\label{table:testSmellDescription}
\begin{tabular}{l|l|p{10cm}|r|r} \toprule
\rowcolor{gray20} \multicolumn{1}{c}{\textbf{Acronym}} & \multicolumn{1}{c}{\textbf{Test Smell}} & \multicolumn{1}{c}{\textbf{Description}} & \multicolumn{1}{c}{\textbf{Precision}} & \multicolumn{1}{c}{\textbf{Recall}} \\ \midrule
\rowcolor{gray10} AR &	Assertion Roulette	&	A test method contains assertion statements without an explanation/message	&	94.7\%	&	90.0\% \\
DA	&	Duplicate Assert	&	A test method that contains more than one assertion statement with the same parameters	&	85.7\%	&	90.0\% \\
\rowcolor{gray10} ECT	&	Handling Exception	&	A test method that contains throws statements	&	100.0\%	&	100.0\%	\\
ET	&	Eager Test	&	A test method contains multiple calls to multiple production methods	&	100.0\%	&	100.0\%	\\
\rowcolor{gray10} GF	&	General Fixture	&	Fields within the setUp method are not utilized by all test methods		&	95.2\%	&	100.0\%	\\
LT & Lazy Test & Multiple test methods call the same class under test methods & 90.9\% & 100.0\% \\
\midrule
\end{tabular}
\end{table*}

\begin{table*}[!t]
\centering \footnotesize
\caption{\revised{Description of refactorings detected by \textsc{TestRefactoringMiner} tool}}
\label{table:refactorings}
\setlength{\tabcolsep}{0.2cm}
\begin{tabular}{p{2.6cm}|p{12cm}|r|r}
\toprule

\rowcolor{gray20}\multicolumn{1}{c}{\textbf{Refactoring}} & \multicolumn{1}{c}{\textbf{Description}} &\textbf{Precision} & \textbf{Recall}  \\ \midrule
\rowcolor{gray10} Add assert explanation & Add an optional parameter into the assert methods to provide an explanatory message & 100.0\% & 78.0\%\\
Extract Class & Create a new class and place the fields and methods responsible for the relevant functionality in it & 100.0\% & 100.0\%\\
\rowcolor{gray10}Extract Method & Move a code fragment to a separate new method and replace the old code with a call to the method & 99.9\% & 96.9\%\\
Inline Method & Replace calls to the method with the method’s content and delete the method itself & 100.0\% & 98.2\% \\
\rowcolor{gray10}Parameterize Test & Remove duplicate code using the @parameterized test annotation to define a variety of arguments & 100.0\% & 100.0\%\\
Replace @Test annotation w/ assertThrows & Remove @Test annotation and add of assertThrows method & 100.0\% & 93.0\% \\
\rowcolor{gray10}Replace @Rule annotation w/ assertThrows & Remove @Rule annotation and add of assertThrows method & 100.0\% & 88.0\% \\
Replace try/catch w/ assertThrows & Remove try/catch blocks and add of assertThrows method & 100.0\% & 89.0\% \\
\rowcolor{gray10}Split method & Separate a long function by splitting it into short methods and adding a call for the new methods & 100.00\% & 100.00\% \\ \midrule 
\end{tabular}
\end{table*}

\smallskip
\textbf{Software Systems.} The selection of suitable software systems will be driven by various considerations. First, we will focus on open-source projects, as we need access to change history information. Second, we will rely on popular, large real-world projects having enough releases to collect data that can be analyzed statistically. \revised{Third, we will standardize the building process to ease dependency management and streamline build configurations across all projects. As such, we plan to use SEART tool\footnote{\url{https://seart-ghs.si.usi.ch/}} to select 100 open-source, non-fork projects from \textsc{GitHub} that have at least 100 stars, 10 major releases, 1,000 lines of code, and 10 test classes. We will seek \textsc{Java} projects that can be compiled with Maven and Java 8---Java 8 is the most popular Java version used nowadays.\footnote{ \url{https://www.jetbrains.com/lp/devecosystem-2021/java/}}}
\revised{Should our search identify more than 100 projects, we will apply random sampling and verify whether the projects were properly built until we have 100 projects.}

\revised{It is worth noting that some projects may adopt the so-called \emph{Boy Scout} rule, i.e., ``Leave every piece of code you touch cleaner than how they found it''.\footnote{ The Boy Scout Rule: \url{https://www.oreilly.com/library/view/97-things-every/9780596809515/ch08.html}.} These projects may be more inclined to the application of refactoring and therefore we may observe a higher test code quality and effectiveness. As part of our study, we will manually analyze the contribution guidelines of the projects selected, looking for any insight suggesting that those projects follow the \emph{Boy Scout} rule. Should we identify a decent amount of projects, we will perform an additional analysis, comparing the results obtained between Boy Scout and non-Boy Scout projects.}

\smallskip
\textbf{Empirical Study Variables.} In the context of \textbf{RQ$_1$} and \textbf{RQ$_2$}, we are interested in assessing whether refactoring operations are more likely to be observed on test classes exhibiting test code quality and effectiveness concerns. As such, we define the following empirical study variables:

\begin{description}[leftmargin=0.3cm]

    \item \emph{Independent Variables.} These are the factors that will be related to the application of test refactoring, namely (i) test code quality metrics; (ii) presence of test smells (of different types); (iii) branch coverage; and (iv) mutation coverage. Tables \ref{table:metricsDescription} and \ref{table:testSmellDescription} list and describe the independent variables of the study. These metrics will be all computed across releases of different software systems and will be statistically analyzed as described later in this section. The selection of these independent variables is driven by multiple considerations. First, we consider test code quality metrics and test smells that were targeted by previous research in the field \cite{catolino2019experience,
    pecorelli2021relation} and found to impact test code in different manners \cite{Spadini2018_Change_Proneness,Kim2021_Secret_Life}. Second, branch and mutation coverage are widely considered as two key indicators of test code effectiveness, which may estimate the goodness of test cases in dealing with real defects \cite{kochhar2017code,papadakis2018mutation}. 

    \smallskip
    \item \emph{Dependent Variables.} These are the refactoring operations (of different types) being observed across releases of different software systems. To select suitable test refactoring operations for our purpose, we investigated the literature to elicit the test refactoring operations that were previously associated to our independent variables---we basically surveyed the previous papers on the matter, discussed in Section \ref{section:Related_Work}, to extract the test refactoring types that researchers have been observing as potentially impacting on testing evolutionary activities. Table \ref{table:refactorings} lists the refactoring operations that will be targeted, along with a brief description. 

\end{description}

When it turns to \textbf{RQ$_3$} and \textbf{RQ$_4$}, we are interested in assessing the impact of test refactoring on the test code quality and effectiveness aspects considered. As such, we need to swap independent and dependent variables: indeed, in this case we are interested to observe how refactoring impacts test code properties rather than the opposite:

\begin{description}[leftmargin=0.3cm]

    \item \emph{Independent Variables.} These are the different types of refactoring operations (Table \ref{table:refactorings}) computed across the releases of software systems considered.

    \smallskip
    \item \emph{Dependent Variables.} These will be the test code quality and effectiveness metrics described in Tables \ref{table:metricsDescription} and \ref{table:testSmellDescription}, which will be computed across releases of software systems.

\end{description}

\revised{In all \textbf{RQ}s we will include a number of control variables, which will help us better verify the extent to which test refactoring impacts the variation of test code quality and effectiveness in relation to project- and process-level characteristics that may impact the dependent variables.}

\begin{description}[leftmargin=0.3cm]
    \item \revised{\emph{Control Variables.} We will first account for the frequency of releases and activities by the project, as these may provide insights into the development speed which, in turn, may impact test code quality and effectiveness. Given a release $R_i$, we will compute the number of releases issued within the last 1, 3, 6, and 12 months. In addition, for each class $C_j$ within $R_i$, we will compute the number of commits performed by developers between the releases $R_{i-1}$ and $R_i$. We will also consider project-level metrics such as (1) project size in terms of lines of code; (2) number of contributors; (3) number of branches; and (4) number of pull requests. On the one hand, these metrics can well overview the main characteristics of the project and the community around it. On the other hand, all these metrics can impact in various manners test code quality and effectiveness, e.g., a higher amount of branches may indicate a higher level of activity around the project, which in turn can influence the way test cases are maintained and evolved.}
\end{description}

\smallskip
\subsection{Data Collection}
\label{subsection:Experimental_environment}
We will use different automated tools available in the literature to extract data on quality and effectiveness metrics, test smells, and refactoring operations. Then, we will merge the data to compose our dataset.

\smallskip
\textbf{Collecting test code quality and effectiveness metrics.}
To collect both test code quality and effectiveness metrics (Table \ref{table:metricsDescription}), we will run \textsc{VITRuM}, a plug-in for the visualization of test-related metrics in order to calculate five static metrics and three dynamic metrics from the test code \cite{Pecorelli2020_Vitrum}. Note that the tool uses \textsc{JaCoCo} to calculate line and branch coverage, and \textsc{pitest} for the mutation coverage. Therefore, we will have to build the projects to calculate the dynamic metrics. 

\smallskip
\textbf{Collecting test smells.} 
Among the test smell detection tools available for \textsc{Java} code \cite{Aljedaani21_Mapping_Tools}, we will use \textsc{tsDetect} \cite{Peruma2020tsdetect}, which is the most accurate tool, with a precision score ranging from 85\% to 100\% and a recall score ranging from 90\% to 100\%. 
\textsc{tsDetect} performs a test code static analysis through an AST (Abstract Syntax Tree) to apply the test smells detection rules in the test files. A test file in the JUnit testing framework should follow the naming conventions of either pre-pending or appending the word \texttt{``Test''} to the name of the production class under test and at the same package hierarchy \cite{Peruma2020tsdetect}. With the detection rules, the tool can detect (i) the presence or absence of a test smell in a test class, or (ii) the number of instances per test smell in a test class. In addition, the tool receives a configuration of the severity thresholds for each test smell \cite{Spadini2020_Severity}. We run the tool to identify the number of instances of the six test smells described in Table \ref{table:testSmellDescription} with default values for the severity thresholds (i.e., the tool reports all instances of test smells detected). 

\smallskip
\textbf{Collecting refactoring data.} To detect test refactoring operations, we will use the \textsc{TestRefactoringMiner} tool \cite{martins2023Refactoring}. 
The tool is built on top of the state-of-the-art refactoring mining tool \textsc{RefactoringMiner}, which has the highest precision (99.8\%) and recall (97.6\%) scores among the currently available refactoring mining tools \cite{RefactoringMiner2022}. In more detail, \textsc{TestRefactoringMiner} analyzes the added, deleted, and changed files between two project versions to detect specific test refactorings, reaching 100\% and 92.5\% of precision and recall scores. The tool operationalizes the detection of all the refactoring operations considered in the study---\revised{Table \ref{table:refactorings} presents the set of test refactoring that we will investigate in this study. It is worth noting that this set considers various refactoring operations, such as integrating new technologies like JUnit 5 or improving the organization of test classes.}



\smallskip
\textbf{Data integration.} Although some tools allow a finer granularity during the code analysis, all of them can also report the results at the class level. Therefore, we will establish traceability links between the test classes reported by \textsc{tsDetect}, \textsc{VITRuM}, and \textsc{TestRefactoringMiner} tools, finally integrating their outcome in a unique data source to be further analyzed from a statistical standpoint. 

\subsection{Data Analysis}
\label{subsection:Data_Analysis}
We first formulate the working hypotheses that we will later statistically assess. As for \textbf{RQ$_1$}, given a quality metric $Qm_i$, with $Qm_i$ in \{LOC, NOM, WMC, RFC, AD\} and a refactoring $ref_k$ in the set of refactoring operations considered in the study, our null hypothesis is the following:

\begin{description}

    \item[\textbf{Hn1$_{Qm_i-ref_k}$.}] There is \emph{no significant difference} in terms of the amount of $ref_k$ operations performed on test classes having different values of $Qm_i$.
    
\end{description}

As in \textbf{RQ$_1$}, we will also evaluate the relation between test refactoring and test smells. Given a test smell $Ts_i$ in the set of test smells considered in the study and $ref_k$, we define a second null hypothesis:

\begin{description}

    \item[\textbf{Hn2$_{Ts_i-ref_k}$.}] There is \emph{no significant difference} in terms of the amount of $ref_k$ operations performed on test classes affected and not by $Ts_i$.
    
\end{description}

As for \textbf{RQ$_2$}, given an effectiveness metric $Em_i$, where $Em_i$ assumes values in the set \{Branch Coverage and Mutation Coverage\} and $ref_k$, the null hypothesis is the following:

\begin{description}

    \item[\textbf{Hn3$_{Em_i-ref_k}$.}] There is \emph{no significant difference} in terms of the amount of $ref_k$ operations performed on test classes having different values of $Em_i$.
    
\end{description}

As for \textbf{RQ$_3$}, given a quality metric $Qm_i$, a test smell $Ts_i$, and a refactoring $ref_k$, the null hypotheses is:

\begin{description}

    \item[\textbf{Hn4$_{Qm_i-ref_k}$.}] There is \emph{no significant difference} in terms of $Qm_i$ before and after the application of $ref_k$.

    \smallskip
    \item[\textbf{Hn5$_{Ts_i-ref_k}$.}] There is \emph{no significant difference} in the number of $Ts_i$ instances before and after the application of $ref_k$.
    
\end{description}

Finally, as for \textbf{RQ$_4$}, the null hypothesis will be:

\begin{description}

    \item[\textbf{Hn6$_{Em_i-ref_k}$.}] There is \emph{no significant difference} in terms of $Em_i$ before and after the application of $ref_k$.

\end{description}

If one of the null hypotheses will be statistically rejected, we will accept the corresponding alternative hypothesis, namely:

\begin{description}

    \item[\textbf{An1$_{Qm_i-ref_k}$.}] The amount of $ref_k$ operations on test classes having different values of $Qm_i$ is \emph{statistically different}.

    \smallskip
    \item[\textbf{An2$_{Ts_i-ref_k}$.}] The amount of $ref_k$ operations on test classes affected and not by $Ts_i$ is \emph{statistically different}.

    \smallskip
    \item[\textbf{An3$_{Em_i-ref_k}$.}] The amount of $ref_k$ operations on test classes having different values of $Em_i$ is \emph{statistically different}.

    \smallskip
    \item[\textbf{An4$_{Qm_i-ref_k}$.}] The $Qm_i$ before and after the application of $ref_k$ is \emph{statistically different}.

    \smallskip
    \item[\textbf{An5$_{Ts_i-ref_k}$.}] The number of $Ts_i$ instances before and after the application of $ref_k$ is \emph{statistically different}. 
    
    \smallskip
    \item[\textbf{An6$_{Em_i-ref_k}$.}] The $Em_i$ before and after the application of $ref_k$ is \emph{statistically different}. 
   
\end{description}

We will then verify the working hypotheses, hence accepting or rejecting them, by building statistical models.

\smallskip
\textbf{Statistical modeling for RQ$_1$ and RQ$_2$.} To address our first two research questions, we will devise a \emph{Logistic Regression Model} for each refactoring operation considered in the study. Such a model belongs to the class of Generalized Linear Models (GLM) \cite{nelder1972generalized} and relates a (dichotomous) dependent variable---in our case, whether or not a particular type of refactoring is performed---with either continuous and discrete independent variables---the quality and effectiveness metrics considered in \textbf{RQ$_1$} and \textbf{RQ$_2$}. 

Before building the statistical model, we plan to assess the presence of multi-collinearity \cite{o2007caution}, which arises when two or more independent variables are highly correlated and can be predicted one from the other. We will use the \texttt{vif} (Variance Inflation Factors) function and discard highly correlated variables, putting a threshold value equal to 5 \cite{o2007caution}. 

For each statistical model, we assess (i) whether each independent variable is significantly correlated with the dependent variable (using a significance level of $\alpha$ = 5\%, and (ii) quantify this correlation using the Odds Ratio (OR) \cite{bland2000odds}, which is a measure of the strength of the association between each independent variable and the dependent variable. Higher OR values for an independent variable indicate a higher probability of explaining the dependent variable, i.e., a higher likelihood that a refactoring operation has been triggered by the independent variable. Nonetheless, the interpretation of OR values change depending on the different measurement scale of the independent variables, i.e., ratio for the test code quality and effectiveness metrics and categorical for the test smells. As for the metrics, the OR for an independent variable indicates the increment of chances for a test class to be subject of refactoring as a consequence of a one-unit increase of the independent variable. As for test smells, the OR indicates how likely a smelly test class is involved in refactoring operations with respect to a non-affected class.

The statistical significance of the correlation between independent and dependent variables will allow us to accept or reject \textbf{Hn1$_{Qm_i-ref_k}$}, \textbf{Hn2$_{Ts_i-ref_k}$}, and \textbf{Hn3$_{Em_i-ref_k}$}, while OR values will measure the strengths of the correlations.


\smallskip
\textbf{Statistical modeling for RQ$_3$ and RQ$_4$.} To statistically assess the impact of test refactoring on test code quality and effectiveness metrics and smells, we will first collect all the test classes subject to the refactoring type $ref_k$ in a generic release $R_i$. Afterward, for each of those test classes, we will compute its value of test code quality and effectiveness metrics and smells computed on the release $R_i$ and the value of the metrics and smells computed on the release $R_{i-1}$. 

We will produce two distributions: the first representing the metric values (or the number of test smells) in $R_{i-1}$, i.e., before the application of $ref_k$; the second representing the metric values (or the number of test smells) in $R_i$, i.e., after the application of $ref_k$. On this basis, we will employ the non-parametric Wilcoxon Rank Sum Test \cite{mcknight2010mann} (with $\alpha$-value = 0.05), through which we will accept or reject the null hypotheses \textbf{Hn4$_{Qm_i-ref_k}$}, \textbf{Hn5$_{Ts_i-ref_k}$}, and \textbf{Hn6$_{Em_i-ref_k}$}.

In addition, we will also rely on the Vargha-Delaney ($\hat{A}_{12}$)~\cite{vargha2000critique} statistical test to measure the magnitude of the differences observed in the considered distributions. According to the direction and value given by $\hat{A}_{12}$ we will have a practical interpretation of our findings, which will depend on the test code factor considered. Specifically, should the $\hat{A}_{12}$ values be lower than 0.5, this would imply that:

\begin{itemize}
    \item The metric values computed on the release $R_{i-1}$ are lower than those on $R_i$, i.e., the refactoring $ref_k$ would have a \emph{positive} effect on the quality or effectiveness metric considered---lower metric values in $R_{i-1}$ would indeed indicate that the refactoring induced the increase of the metric in $R_i$, hence having a positive effect;

    \smallskip
    \item The number of test smells computed on the release $R_{i-1}$ is lower than the one computed on $R_i$, i.e., the refactoring $ref_k$ would have a \emph{negative} effect, hence suggesting that, rather than improving test code design, the refactoring induced the emergence of some form of test smells. 
\end{itemize}

Similarly, a $\hat{A}_{12} > 0.50$ indicates the opposite, hence that either $ref_k$ has a \emph{negative} impact on the considered test code quality or effectiveness metric, or that the refactoring has a \emph{positive} impact of the removal of test smells. Finally, $\hat{A}_{12} == 0.50$ points out that the results are identical, i.e., the refactoring has limited to no effect on the dependent variables.



\subsection{Publication of generated dataset}
\label{subsection:Publication_dataset}
The dataset that we will collect by merging test code metrics, test smells, test effectiveness metrics, and test refactoring data will be made publicly available in an online repository \revised{\cite{Our_repo}}. We also plan to release the scripts for the data collection and analysis that we will use to perform this study.

\section{Threats to validity}
\label{section:ThreatsToValidity}
This section discusses the potential threats that may affect the validity of our empirical study plan.

\smallskip
\textbf{Construct validity.} \revised{A first threat concerns with the criteria we will use to select software projects: despite the actions to standardize the building process, we might still fall into build failures. Should this happen, we will attempt to manually diagnose the reasons of the failures, trying to fix them - in this respect, we will exploit recent research \cite{tufano2017there, maes2022revisiting} reporting insights on how to fix build failures. In the best case, we would still be able to build the project. In the worst case, we would not be able to fix the build failure and, in this case, we will finally discard the project from our study and replace it with another project retrieved by using the SEART tool.}

As for the set of test smells, structural and dynamic metrics we will use to assess the test code quality. We will not calculate all the Chidamber \& Kemerer metrics as some of them do not apply to the context of test code (e.g., \textit{Depth Inheritance Tree}). Nevertheless, we have chosen a mix of metrics capturing the test code size, structural, and dynamic characteristics. Another threat to validity concerns the identification of test smells and refactoring operations. We will use tools already validated and used by the research community. Although the tools present high precision and recall scores, they might report some false positive or false negative instances of test smells or refactorings: in response of this limitation, we will attempt to perform preliminary, manual investigations to assess the degree of accuracy of the tools before running them on a large scale---in this way, we will be able to provide indications on the confidence level of our conclusions.

\smallskip
\textbf{Internal Validity.} This category of threats to validity concerns by-product changes of other maintenance activities (e.g., bug fixes or changes in requirements) that could also contribute to the removal of test smells. Therefore, the data analysis will not indicate a causal relationship, but rather that there is a possibility of a relationship that may be further investigated. We will attempt to corroborate our quantitative results by means of some qualitative insights. \revised{In addition, we acknowledge test flakiness as a potential threat to the internal validity which can impact the reliability of our findings. However, despite being a severe issue for practitioners, previous investigations found test flakiness to arise in a limited amount of cases, e.g., Luo et al. \cite{luo2014empirical} found out that flaky tests affect up to 4.56\% of test cases. In this sense, it is reasonable to believe that the problem of test flakiness will have a limited impact on our findings.}

\smallskip
\textbf{External Validity.} This class of threats to validity mainly concerns the subject projects of our study. \revised{We selected open-source \textsc{Java} projects from \textsc{GitHub}, which are only a fraction of the complete picture of open-source software and do not necessarily represent industrial practices. Therefore, the results may not generalize to the industrial context and other programming languages. In addition, we will select projects based on the number of stars, which may raise some popularity bias. Replications of our work would be, therefore, beneficial to corroborate our findings in different contexts: to stimulate further research, we will release all materials and scripts as part of an archived online appendix \cite{Our_repo}.} 


\smallskip
\textbf{Conclusion validity.} 
To address how frequently test refactoring is performed on test classes affected by quality or effectiveness concerns, we will use logistic regression models to identify correlations. Other than highlighting cases of significant correlations, we will report and discuss OR values. In addition, to investigate the effect of test refactoring on test code quality and effectiveness, we will employ well-established statistical tests such as the Wilcoxon Rank Sum Test \cite{mcknight2010mann} and the Vargha-Delaney ($\hat{A}_{12}$)~\cite{vargha2000critique} statistical tests. \revised{Our analysis will be conducted at the granularity of classes because the tools we plan to employ work at this level. This may bias our conclusions, as this granularity may be subject to various confounding variables. On the one hand, this is a limitation that we unfortunately share with all the other research works that analyze dynamic test code metrics \cite{kumar2010survey}. On the other hand, we plan for the inclusion of multiple process- and project-level control variables, through which we will be able to partially mitigate this threat to validity.}

\revised{An additional point to remark is that our data collection procedure cannot distinguish between changes that were meant as refactoring and other changes where refactoring was applied as part of other modifications. 
We might have mitigated this limitation by extracting refactoring changes through the analysis of issues and pull requests, i.e., collecting changes explicitly intended as refactoring. Nonetheless, such an alternative method could have biased even further the conclusions drawn for two reasons connected to the availability and reliability of the information available within the developers’ discussions on \textsc{GitHub}. More particularly:}

\begin{description}[leftmargin=0.3cm]
    \item \revised{\emph{Availability.} Previous studies established that developers perform ``floss refactoring'', combining refactoring operations and behavioral change edits within individual commits \cite{murphy2007don}. From a practical standpoint, this means that developers do not often apply refactoring for the sake of refactoring source code, but as an instrument to perform other changes, e.g., to simplify a piece of code before making further evolutionary changes. As such, it is unlikely to find ``pure'' refactoring changes or discussions, in the form of issues or pull requests, around refactoring operations to be applied.}

    \smallskip
    \item \revised{\emph{Reliability.} Literature found that developers not only rarely document refactoring activities explicitly \cite{weissgerber2006identifying, weissgerber2007making}, but also that when they do, they are inconsistent \cite{alomar2021we}, i.e., labeling changes as refactoring, although no refactoring is done at all. Other researchers found out that the term ``refactoring'' is misused, i.e., developers do not often correctly distinguish between refactoring changes and normal code modifications \cite{di2018preliminary}. In this respect, the seminal paper by Murphy-Hill et al. \cite{murphy2011we} reported that \emph{``messages in version histories are unreliable indicators of refactoring activities. This is due to the fact that developers do not consistently report/document refactoring activities''}. This latter observation was also backed up by the findings obtained by Ratzinger et al. \cite{ratzinger2008relation}, who discovered that the extraction of refactoring documentation from repositories may lead to several false positives, as the words used by developers are too generic and do not often refer to real refactoring operations. }
\end{description}

\revised{As a consequence, the analysis of issues and pull requests would have led to unreliable conclusions. On the contrary, the goal of a statistical study is exactly that to identify hidden relations between dependent and independent variables while controlling for possible confounding effects \cite{freedman2009statistical}: we believe that such an approach better fits our research goals. Through a large-scale, statistical investigation, we may indeed end up discovering the intrinsic factors associated with the refactoring actions performed by developers, finally providing evidence of how test refactoring is done in practice. }
\section{Conclusion}
\label{section:Conclusion}
The ultimate goal of our research plan is to understand whether the test code quality and effectiveness provide indications of which test classes are more likely of being refactored and to what extent test refactoring operations can improve the test code quality and effectiveness. We will conduct this study on a set of 100 open-source \textsc{Java} projects, starting from the collection of data on the test code quality, test smells, and refactoring operations arising in the major releases of the projects. Then, we will employ statistical approaches to address the goals of our investigation and, based on the conclusions we will be able to draw, finally provide actionable items and implications for researchers and practitioners. 

As an outcome of our exploratory study, we expect to provide the following key contributions:

\begin{enumerate}
    \item An empirical understanding of the factors triggering test refactoring operations, which comprises an analysis of how test code quality and effectiveness come into play;

    \smallskip
    \item Evidence of the impact of test refactoring on test code quality and effectiveness;

    \smallskip
    \item An online appendix which will provide all material and scripts employed to address the goals of the study.
    
\end{enumerate}

\section*{Acknowledgment}
This study was financed in part by the Coordenação de Aperfeiçoamento de Pessoal de Nível Superior – Brasil (CAPES) – Finance Code 001; and FAPESB grants BOL0188/2020 and PIE0002/2022. Fabio is supported by the Swiss National Science Foundation through the SNF Project No. PZ00P2\_186090 (TED).

\balance
\bibliographystyle{IEEEtran}
\tiny
\bibliography{references}

\begin{thebibliography}{10}
\providecommand{\url}[1]{#1}
\csname url@samestyle\endcsname
\providecommand{\newblock}{\relax}
\providecommand{\bibinfo}[2]{#2}
\providecommand{\BIBentrySTDinterwordspacing}{\spaceskip=0pt\relax}
\providecommand{\BIBentryALTinterwordstretchfactor}{4}
\providecommand{\BIBentryALTinterwordspacing}{\spaceskip=\fontdimen2\font plus
\BIBentryALTinterwordstretchfactor\fontdimen3\font minus
  \fontdimen4\font\relax}
\providecommand{\BIBforeignlanguage}[2]{{%
\expandafter\ifx\csname l@#1\endcsname\relax
\typeout{** WARNING: IEEEtran.bst: No hyphenation pattern has been}%
\typeout{** loaded for the language `#1'. Using the pattern for}%
\typeout{** the default language instead.}%
\else
\language=\csname l@#1\endcsname
\fi
#2}}
\providecommand{\BIBdecl}{\relax}
\BIBdecl

\bibitem{Fowler18}
M.~Fowler, \emph{Refactoring: improving the design of existing code}.\hskip 1em
  plus 0.5em minus 0.4em\relax USA: Addison-Wesley Longman Publishing Co.,
  Inc., 1999.

\bibitem{bavota2014recommending}
G.~Bavota, A.~De~Lucia, A.~Marcus, and R.~Oliveto, ``Recommending refactoring
  operations in large software systems,'' \emph{Recommendation Systems in
  Software Engineering}, pp. 387--419, 2014.

\bibitem{azeem2019machine}
M.~I. Azeem, F.~Palomba, L.~Shi, and Q.~Wang, ``Machine learning techniques for
  code smell detection: A systematic literature review and meta-analysis,''
  \emph{IST}, vol. 108, pp. 115--138, 2019.

\bibitem{de2018systematic}
E.~V. de~Paulo~Sobrinho, A.~De~Lucia, and M.~de~Almeida~Maia, ``A systematic
  literature review on bad smells--5 w's: which, when, what, who, where,''
  \emph{IEEE TSE}, vol.~47, no.~1, pp. 17--66, 2018.

\bibitem{DuBois2004_Refactoring_to_improve_metrics}
B.~Du~Bois, S.~Demeyer, and J.~Verelst, ``Refactoring - improving coupling and
  cohesion of existing code,'' in \emph{11th Working Conf. on Reverse
  Engineering}, 2004, pp. 144--151.

\bibitem{Chavez2017_Refactoring_and_internal_attr}
A.~Ch\'{a}vez, I.~Ferreira, E.~Fernandes, D.~Cedrim, and A.~Garcia, ``How does
  refactoring affect internal quality attributes? a multi-project study,'' in
  \emph{Proceedings of the XXXI Brazilian Symposium on Software Engineering},
  ser. SBES '17.\hskip 1em plus 0.5em minus 0.4em\relax New York, NY, USA:
  Association for Computing Machinery, 2017, p. 74–83.

\bibitem{al2015identifying}
J.~Al~Dallal, ``Identifying refactoring opportunities in object-oriented code:
  A systematic literature review,'' \emph{IST}, vol.~58, pp. 231--249, 2015.

\bibitem{baqais2020automatic}
A.~A.~B. Baqais and M.~Alshayeb, ``Automatic software refactoring: a systematic
  literature review,'' \emph{Software Quality Journal}, vol.~28, no.~2, pp.
  459--502, 2020.

\bibitem{lacerda2020code}
G.~Lacerda, F.~Petrillo, M.~Pimenta, and Y.~G. Gu{\'e}h{\'e}neuc, ``Code smells
  and refactoring: A tertiary systematic review of challenges and
  observations,'' \emph{JSS}, vol. 167, p. 110610, 2020.

\bibitem{di2020relationship}
M.~Di~Penta, G.~Bavota, and F.~Zampetti, ``On the relationship between
  refactoring actions and bugs: a differentiated replication,'' in
  \emph{Proceedings of the 28th ACM Joint Meeting on European Software
  Engineering Conference and Symposium on the Foundations of Software
  Engineering}, 2020, pp. 556--567.

\bibitem{bavota2012does}
G.~Bavota, B.~De~Carluccio, A.~De~Lucia, M.~Di~Penta, R.~Oliveto, and
  O.~Strollo, ``When does a refactoring induce bugs? an empirical study,'' in
  \emph{2012 IEEE 12th International Working Conference on Source Code Analysis
  and Manipulation}.\hskip 1em plus 0.5em minus 0.4em\relax IEEE, 2012, pp.
  104--113.

\bibitem{ferreira2018}
\BIBentryALTinterwordspacing
I.~Ferreira, E.~Fernandes, D.~Cedrim, A.~Uch\^{o}a, A.~C. Bibiano, A.~Garcia,
  J.~a.~L. Correia, F.~Santos, G.~Nunes, C.~Barbosa, B.~Fonseca, and
  R.~de~Mello, ``The buggy side of code refactoring: Understanding the
  relationship between refactorings and bugs,'' in \emph{Proceedings of the
  40th International Conference on Software Engineering: Companion
  Proceeedings}, ser. ICSE '18.\hskip 1em plus 0.5em minus 0.4em\relax New
  York, NY, USA: Association for Computing Machinery, 2018, p. 406–407.
  [Online]. Available: \url{https://doi.org/10.1145/3183440.3195030}
\BIBentrySTDinterwordspacing

\bibitem{iannone2023rubbing}
E.~Iannone, Z.~Codabux, V.~Lenarduzzi, A.~De~Lucia, and F.~Palomba, ``Rubbing
  salt in the wound? a large-scale investigation into the effects of
  refactoring on security,'' \emph{Empirical Software Engineering}, vol.~28,
  no.~4, p.~89, 2023.

\bibitem{tufano2015and}
M.~Tufano, F.~Palomba, G.~Bavota, R.~Oliveto, M.~Di~Penta, A.~De~Lucia, and
  D.~Poshyvanyk, ``When and why your code starts to smell bad,'' in \emph{2015
  IEEE/ACM 37th IEEE International Conference on Software Engineering},
  vol.~1.\hskip 1em plus 0.5em minus 0.4em\relax IEEE, 2015, pp. 403--414.

\bibitem{murphy2007don}
E.~Murphy-Hill and A.~P. Black, ``Why don’t people use refactoring tools?''
  in \emph{Proceedings of the 1st Workshop on Refactoring Tools}, 2007, pp.
  61--62.

\bibitem{micco2017state}
J.~Micco, ``The state of continuous integration testing@ google,'' 2017.

\bibitem{grano2020pizza}
G.~Grano, C.~De~Iaco, F.~Palomba, and H.~C. Gall, ``Pizza versus pinsa: On the
  perception and measurability of unit test code quality,'' in \emph{2020 IEEE
  Int.l Conf. on Software Maintenance and Evolution (ICSME)}.\hskip 1em plus
  0.5em minus 0.4em\relax IEEE, 2020, pp. 336--347.

\bibitem{meszaros2007xunit}
G.~Meszaros, \emph{xUnit test patterns: Refactoring test code}.\hskip 1em plus
  0.5em minus 0.4em\relax Pearson Education, 2007.

\bibitem{Meszaros2003_Manifesto}
G.~Meszaros, S.~M. Smith, and J.~Andrea, ``The test automation manifesto,''
  ser. Extreme Programming and Agile Methods - XP/Agile Universe 2003,
  F.~Maurer and D.~Wells, Eds., Springer Berlin Heidelberg.\hskip 1em plus
  0.5em minus 0.4em\relax Berlin, Heidelberg: Springer, 2003, pp. 73--81.

\bibitem{Guerra2007RefactoringSafely}
E.~M. Guerra and C.~T. Fernandes, ``Refactoring test code safely,'' in
  \emph{Int.l Conf. on Software Engineering Advances (ICSEA 2007)}.\hskip 1em
  plus 0.5em minus 0.4em\relax New York, NY, USA: IEEE, 2007, pp. 44--44.

\bibitem{Deursen01_Refactoring}
A.~Deursen, L.~M. Moonen, A.~Bergh, and G.~Kok, ``Refactoring test code,''
  Centre for Mathematics and Computer Science, NLD, Tech. Rep., 2001.

\bibitem{Soares2020_Refactoring_Smells}
E.~Soares, M.~Ribeiro, G.~Amaral, R.~Gheyi, L.~Fernandes, A.~Garcia,
  B.~Fonseca, and A.~Santos, ``Refactoring test smells: A perspective from
  open-source developers,'' in \emph{Proceedings of the 5th Brazilian Symposium
  on Systematic and Automated Software Testing}, ser. SAST 20.\hskip 1em plus
  0.5em minus 0.4em\relax New York, NY, USA: Association for Computing
  Machinery, 2020, p. 50–59.

\bibitem{Soares22_JUnit5}
E.~Soares, M.~Ribeiro, R.~Gheyi, G.~Amaral, and A.~M. Santos, ``Refactoring
  test smells with junit 5: Why should developers keep up-to-date,'' \emph{IEEE
  TSE}, pp. 1--1, 2022.

\bibitem{Peruma2020_Refactoring_Tests_Prod}
A.~Peruma, C.~D. Newman, M.~W. Mkaouer, A.~Ouni, and F.~Palomba, ``An
  exploratory study on the refactoring of unit test files in android
  applications,'' in \emph{Proceedings of the 42nd Int.l Conf. on Software
  Engineering Workshops}, ser. ICSEW'20.\hskip 1em plus 0.5em minus 0.4em\relax
  New York, NY, USA: Association for Computing Machinery, 2020, p. 350–357.

\bibitem{Spadini2018_CodeQuality}
D.~Spadini, F.~Palomba, A.~Zaidman, M.~Bruntink, and A.~Bacchelli, ``On the
  relation of test smells to software code quality,'' in \emph{2018 IEEE Int.l
  Conf. on Software Maintenance and Evolution (ICSME)}.\hskip 1em plus 0.5em
  minus 0.4em\relax New York, NY, USA: IEEE, 2018, pp. 1--12.

\bibitem{Haitao2022_ImproveQuality}
H.~Wu, R.~Yin, J.~Gao, Z.~Huang, and H.~Huang, ``To what extent can code
  quality be improved by eliminating test smells?'' in \emph{2022 Int.l Conf.
  on Code Quality (ICCQ)}.\hskip 1em plus 0.5em minus 0.4em\relax New York, NY,
  USA: IEEE, 2022, pp. 19--26.

\bibitem{Kim2021_Secret_Life}
D.~J. Kim, T.-H.~P. Chen, and J.~Yang, ``The secret life of test smells-an
  empirical study on test smell evolution and maintenance,'' \emph{EMSE},
  vol.~26, no.~5, pp. 1--47, 2021.

\bibitem{RefactoringMiner2022}
N.~Tsantalis, A.~Ketkar, and D.~Dig, ``Refactoringminer 2.0,'' \emph{IEEE TSE},
  vol.~48, no.~3, pp. 930--950, 2022.

\bibitem{Peruma2020tsdetect}
A.~Peruma, K.~Almalki, C.~D. Newman, M.~W. Mkaouer, A.~Ouni, and F.~Palomba,
  ``tsdetect: an open source test smells detection tool,'' ser. Symposium on
  the Foundations of Software Engineering, Association for Computing
  Machinery.\hskip 1em plus 0.5em minus 0.4em\relax ACM, 2020.

\bibitem{Damasceno2023_Analyzing_Refactorings}
H.~Damasceno, C.~Bezerra, E.~Coutinho, and I.~Machado, ``Analyzing test smells
  refactoring from a developers perspective,'' in \emph{Proceedings of the XXI
  Brazilian Symposium on Software Quality}, ser. SBQS '22.\hskip 1em plus 0.5em
  minus 0.4em\relax New York, NY, USA: Association for Computing Machinery,
  2023.

\bibitem{Wohlin2012_Experimentation}
C.~Wohlin, P.~Runeson, M.~H{\"o}st, M.~C. Ohlsson, B.~Regnell, and
  A.~Wessl{\'e}n, \emph{Experimentation in software engineering}.\hskip 1em
  plus 0.5em minus 0.4em\relax Springer Science \& Business Media, 2012.

\bibitem{Pecorelli2020_Vitrum}
F.~Pecorelli, G.~Di~Lillo, F.~Palomba, and A.~De~Lucia, ``Vitrum: A plug-in for
  the visualization of test-related metrics,'' in \emph{Proceedings of the
  Int.l Conf. on Advanced Visual Interfaces}, ser. AVI'20.\hskip 1em plus 0.5em
  minus 0.4em\relax New York, NY, USA: Association for Computing Machinery,
  2020.

\bibitem{catolino2019experience}
G.~Catolino, F.~Palomba, A.~Zaidman, and F.~Ferrucci, ``How the experience of
  development teams relates to assertion density of test classes,'' in
  \emph{2019 IEEE Int.l Conf. on Software Maintenance and Evolution
  (ICSME)}.\hskip 1em plus 0.5em minus 0.4em\relax IEEE, 2019, pp. 223--234.

\bibitem{pecorelli2021relation}
F.~Pecorelli, F.~Palomba, and A.~De~Lucia, ``The relation of test-related
  factors to software quality: a case study on apache systems,'' \emph{EMSE},
  vol.~26, pp. 1--42, 2021.

\bibitem{Spadini2018_Change_Proneness}
D.~Spadini, F.~Palomba, A.~Zaidman, M.~Bruntink, and A.~Bacchelli, ``On the
  relation of test smells to software code quality,'' in \emph{2018 IEEE Int.l
  Conf. on Software Maintenance and Evolution (ICSME)}, 2018, pp. 1--12.

\bibitem{kochhar2017code}
P.~S. Kochhar, D.~Lo, J.~Lawall, and N.~Nagappan, ``Code coverage and
  postrelease defects: A large-scale study on open source projects,''
  \emph{IEEE Transactions on Reliability}, vol.~66, no.~4, pp. 1213--1228,
  2017.

\bibitem{papadakis2018mutation}
M.~Papadakis, D.~Shin, S.~Yoo, and D.-H. Bae, ``Are mutation scores correlated
  with real fault detection? a large scale empirical study on the relationship
  between mutants and real faults,'' in \emph{Proceedings of the 40th Int.l
  Conf. on Software Engineering}, 2018, pp. 537--548.

\bibitem{Aljedaani21_Mapping_Tools}
W.~Aljedaani, A.~Peruma, A.~Aljohani, M.~Alotaibi, M.~W. Mkaouer, A.~Ouni,
  C.~D. Newman, A.~Ghallab, and S.~Ludi, ``Test smell detection tools: A
  systematic mapping study,'' ser. Evaluation and Assessment in Software
  Engineering.\hskip 1em plus 0.5em minus 0.4em\relax New York, NY, USA:
  Association for Computing Machinery, 2021, p. 170–180.

\bibitem{Spadini2020_Severity}
D.~Spadini, M.~Schvarcbacher, A.-M. Oprescu, M.~Bruntink, and A.~Bacchelli,
  ``Investigating severity thresholds for test smells,'' in \emph{Proceedings
  of the 17th Int.l Conf. on Mining Software Repositories}, ser. MSR '20.\hskip
  1em plus 0.5em minus 0.4em\relax New York, NY, USA: Association for Computing
  Machinery, 2020, p. 311–321.

\bibitem{martins2023Refactoring}
L.~Martins, H.~Costa, M.~Ribeiro, F.~Palomba, and I.~Machado, ``Automating
  test-specific refactoring mining: A mixed-method investigation,'' in
  \emph{Proceedings of the 23rd IEEE International Working Conference on Source
  Code Analysis and Manipulation}, 2023.

\bibitem{nelder1972generalized}
J.~A. Nelder and R.~W. Wedderburn, ``Generalized linear models,'' \emph{Journal
  of the Royal Statistical Society: Series A (General)}, vol. 135, no.~3, pp.
  370--384, 1972.

\bibitem{o2007caution}
R.~M. O’brien, ``A caution regarding rules of thumb for variance inflation
  factors,'' \emph{Quality \& quantity}, vol.~41, pp. 673--690, 2007.

\bibitem{bland2000odds}
J.~M. Bland and D.~G. Altman, ``The odds ratio,'' \emph{Bmj}, vol. 320, no.
  7247, p. 1468, 2000.

\bibitem{mcknight2010mann}
P.~E. McKnight and J.~Najab, ``Mann-whitney u test,'' \emph{The Corsini
  encyclopedia of psychology}, pp. 1--1, 2010.

\bibitem{vargha2000critique}
A.~Vargha and H.~D. Delaney, ``A critique and improvement of the cl common
  language effect size statistics of mcgraw and wong,'' \emph{Journal of
  Educational and Behavioral Statistics}, vol.~25, no.~2, pp. 101--132, 2000.

\bibitem{Our_repo}
\BIBentryALTinterwordspacing
``Data collection and analysis,'' 2023, {A}ccessed on 12.07.2023. [Online].
  Available: \url{https://figshare.com/s/dd0730e0036ebe4f878b}
\BIBentrySTDinterwordspacing

\bibitem{tufano2017there}
M.~Tufano, F.~Palomba, G.~Bavota, M.~Di~Penta, R.~Oliveto, A.~De~Lucia, and
  D.~Poshyvanyk, ``There and back again: Can you compile that snapshot?''
  \emph{Journal of Software: Evolution and Process}, vol.~29, no.~4, p. e1838,
  2017.

\bibitem{maes2022revisiting}
M.~Maes-Bermejo, M.~Gallego, F.~Gort{\'a}zar, G.~Robles, and J.~M.
  Gonzalez-Barahona, ``Revisiting the building of past snapshots—a
  replication and reproduction study,'' \emph{EMSE}, vol.~27, no.~3, p.~65,
  2022.

\bibitem{luo2014empirical}
Q.~Luo, F.~Hariri, L.~Eloussi, and D.~Marinov, ``An empirical analysis of flaky
  tests,'' in \emph{Proceedings of the 22nd ACM SIGSOFT Int.l Symposium on
  Foundations of Software Engineering}, 2014, pp. 643--653.

\bibitem{kumar2010survey}
J.~Kumar~Chhabra and V.~Gupta, ``A survey of dynamic software metrics,''
  \emph{Journal of computer science and technology}, vol.~25, pp. 1016--1029,
  2010.

\bibitem{weissgerber2006identifying}
P.~Wei{\ss}gerber and S.~Diehl, ``Identifying refactorings from source-code
  changes,'' in \emph{21st IEEE/ACM international conference on automated
  software engineering (ASE'06)}.\hskip 1em plus 0.5em minus 0.4em\relax IEEE,
  2006, pp. 231--240.

\bibitem{weissgerber2007making}
P.~Wei{\ss}gerber, B.~Biegel, and S.~Diehl, ``Making programmers aware of
  refactorings.'' in \emph{WRT}, 2007, pp. 58--59.

\bibitem{alomar2021we}
E.~A. AlOmar, A.~Peruma, M.~W. Mkaouer, C.~Newman, A.~Ouni, and M.~Kessentini,
  ``How we refactor and how we document it? on the use of supervised machine
  learning algorithms to classify refactoring documentation,'' \emph{Expert
  Systems with Applications}, vol. 167, p. 114176, 2021.

\bibitem{di2018preliminary}
Z.~Di, B.~Li, Z.~Li, and P.~Liang, ``A preliminary investigation of
  self-admitted refactorings in open source software (s),'' in \emph{Int.l
  Conf. on Software Engineering and Knowledge Engineering}, vol. 2018.\hskip
  1em plus 0.5em minus 0.4em\relax KSI Research Inc. and Knowledge Systems
  Institute Graduate School, 2018, pp. 165--168.

\bibitem{murphy2011we}
E.~Murphy-Hill, C.~Parnin, and A.~P. Black, ``How we refactor, and how we know
  it,'' \emph{IEEE TSE}, vol.~38, no.~1, pp. 5--18, 2011.

\bibitem{ratzinger2008relation}
J.~Ratzinger, T.~Sigmund, and H.~C. Gall, ``On the relation of refactorings and
  software defect prediction,'' in \emph{Proceedings of the 2008 Int.l working
  Conf. on MSR}, 2008, pp. 35--38.

\bibitem{freedman2009statistical}
D.~A. Freedman, \emph{Statistical models: theory and practice}.\hskip 1em plus
  0.5em minus 0.4em\relax cambridge university press, 2009.

\end{thebibliography}

\end{document}